\documentclass[twocolumn,amsmath,amssymb,aps,prb,floatfix,showpacs,citeautoscript]{revtex4}

\usepackage{graphicx}
\usepackage{dcolumn}
\usepackage{amsmath}
\usepackage{bm}

\begin{document}

\title{Theoretical investigation of polarization-compensated
II-IV/I-V perovskite superlattices}

\author{\'Eamonn D.~Murray}
  \affiliation{Department of Physics and Astronomy, Rutgers University,
Piscataway, New Jersey 08854-8019, USA}
\author{David Vanderbilt}
  \affiliation{Department of Physics and Astronomy, Rutgers University,
Piscataway, New Jersey 08854-8019, USA}

\date{\today}

\begin{abstract}
Recent work suggested that head-to-head and tail-to-tail domain
walls could be induced to form in ferroelectric superlattices by
introducing compensating ``delta doping'' layers via chemical
substitution in specified atomic planes [Phys.\ Rev.\ B {\bf 73},
020103(R), 2006].  Here we investigate a variation of this approach
in which superlattices are formed of alternately stacked groups of
II-IV and I-V perovskite layers, and the ``polar
discontinuity'' at the II-IV/I-V interface effectively provides the
delta-doping layer.  Using first-principles calculations on
SrTiO$_3$/KNbO$_3$ as a model system, we show that this strategy
allows for the growth of a superlattice with stable polarized
regions and large polarization discontinuities at the internal
interfaces.  We also generalize a Wannier-based definition of layer
polarizations in perovskite superlattices [Phys.\ Rev.\ Lett.\ {\bf
97}, 107602 (2006)] to the case in which some (e.g., KO or NbO$_2$)
layers are non-neutral, and apply this method to quantify the local
variations in polarization in the proposed SrTiO$_3$/KNbO$_3$
superlattice system.
\end{abstract}

\maketitle

Ferroelectric materials have been the subject of increasing
theoretical and experimental study in recent years. In particular,
multicomponent superlattices based on ABO$_3$ perovskites have been
shown to possess many interesting properties (see
Ref.~~\onlinecite{ghosezbook} and references therein).
Currently, heterointerfaces between different ABO$_3$ perovskites
are also the subject of intense
investigation\cite{ohtomo04,nakagawa06,cen08,Rijnders08,kalabukhov}.
Given the huge number of possible superlattice arrangements, it is
prudent to investigate systems of potential interest using
first-principles calculations to confirm a given system possesses the
desired properties before experimental work is performed
\cite{George05,Rijnders05}.

In recent work, Wu and Vanderbilt \cite{wuvand06} introduced a
novel concept in which 180$^\circ$ head-to-head (HH) and
tail-to-tail (TT) domain walls are induced to form in a
ferroelectric superlattice via the insertion of compensating
``delta doping'' layers.  For example, in a II-IV ABO$_3$
perovskite, column III or V ions could replace the IV ions in one
BO$_2$ layer, inducing the formation of a HH or TT domain wall
respectively.  Usually the formation of a HH or TT domain wall
perpendicular to the polarization direction would entail an
unacceptable Coulomb energy cost, or cause the domain wall to become
metallic in order that free carriers could compensate the domain
wall. However, it was shown that the delta-doping layers could be
arranged to compensate the polarization bound charge and allow a
structure in which the ferroelectric domains are polarized in
opposite directions and are separated by HH and TT domain walls.

Here we examine the possibility of forming a multicomponent
perovskite superlattice with similarly large discontinuities in the
local electric polarization by making use of alternating II-IV and
I-V perovskite constituents.  In this case, the ``polar
discontinuity'' associated with the II-IV/I-V interface plays the
role of delta-doping layer and compensates the polarization bound
charge at the interface.  (Clearly a similar strategy could be
applied to II-IV/III-III perovskite superlattices.)  The resulting
structure is not switchable, but is locally polarized and has
strongly broken inversion symmetry.  We demonstrate this concept
via first-principles calculations on SrTiO$_3$/KNbO$_3$ as a
prototypical system, showing successful compensation and robust
formation of locally polarized regions.  Superlattices of this type
are shown to remain insulating to rather large layer thicknesses.
Finally, we clarify how the Wannier-based definition of layer
polarization in perovskite superlattices introduced in
Ref.~~\onlinecite{wu06} can be generalized for a system having
non-neutral AO or BO$_2$ constituent layers, and apply this to the
SrTiO$_3$/KNbO$_3$ system to map out the local variations in
polarization.

\begin{figure}
\includegraphics[width=2.8in]{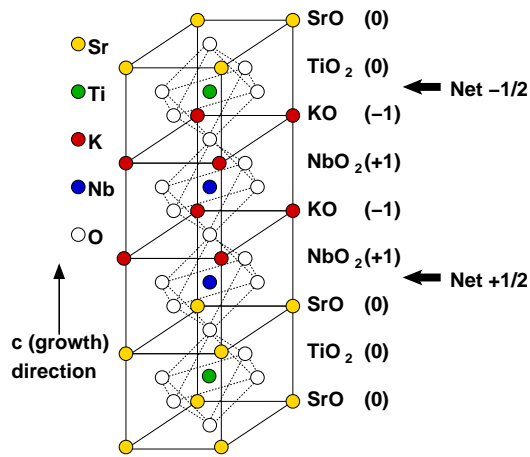}
\caption{\label{fig:supercell} (Color online) 
Sketch of a possible II-IV (SrTiO$_3$) / I-V
(KNbO$_3$) superlattice grown along [001].}
\end{figure}

The type of superlattice structure we have in mind is illustrated
in Fig.~\ref{fig:supercell}. Here, two unit cells of KNbO$_3$ (in
the sequence NbO$_2$--KO--NbO$_2$--KO) repeatedly alternate with
two unit cells of SrTiO$_3$ (in the sequence
TiO$_2$--SrO--TiO$_2$--SrO) during growth.  While each added
KNbO$_3$ or SrTiO$_3$ unit cell is neutral, in KNbO$_3$ this
neutrality results from the cancellation of charges on the KO$^-$
and NbO$_2^+$ layers, while in SrTiO$_3$ the individual layers are
neutral.  When the layers are assembled as in
Fig.~\ref{fig:supercell}, the presence of the ``polar
discontinuity'' introduces effective compositional charges of
$\pm1/2$ at the NbO$_2$/SrO and TiO$_2$/KO interfaces respectively,
as shown.  Similar effects have recently been extensively discussed
for SrTiO$_3$/LaAlO$_3$ and related
interfaces.\cite{ohtomo04,nakagawa06,cen08,Rijnders08} Intuitively,
we may regard each KO$^-$ layer in bulk KNbO$_3$ as being
half-compensated from each of its two immediate NbO$_2^+$
neighbors, whereas at the TiO$_2$/KO interface the KO$^-$ is only
half-compensated because it has only a single NbO$_2^+$ neighbor.
The resulting compositional charge densities are $\sigma_{\rm
comp}=\pm e/2a^2$ at the NbO$_2$/SrO and TiO$_2$/KO interfaces
respectively.

\begin{figure}
 \includegraphics[width=2.5in]{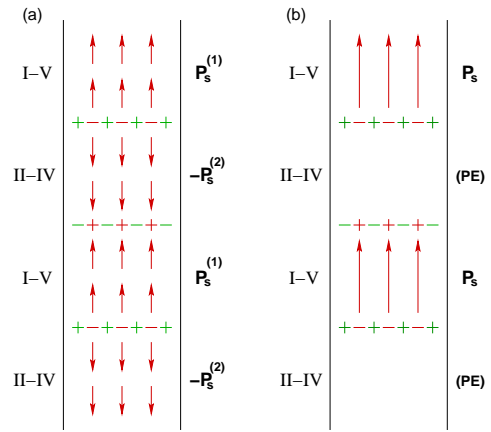}
\caption{\label{fig:layers} (Color online) 
Two possible II-IV/I-V superlattice arrangements
yielding compensating heterointerfaces and stabilized
ferroelectric discontinuities. (a) Both materials are
ferroelectric, with antiparallel polarizations along the growth
direction.  (b) One material is paraelectric and the other
ferroelectric.  Green and red interface charges denote
polar-discontinuity and polarization-induced charges, respectively.  } 
\end{figure}

Two possibilities for cancelling these compositional charges in a
I-V/II-IV superlattice are shown in Fig.~\ref{fig:layers}.  In
Fig.~\ref{fig:layers}(a) we assume that both I-V and II-IV
materials are ferroelectrics, and to take an extreme case we assume
their spontaneous polarizations are equal ($P_{\rm s}^{(1)}=P_{\rm
s}^{(2)}$) at the specified in-plane lattice constant.  Arranging
the polarizations so that they alternate up and down as shown, the
ideal compensation $\sigma_{\rm bound}+\sigma_{\rm comp}=0$ between
polarization bound charge and compositional interface charge is
realized when $P_{\rm s}^{(1)}=P_{\rm s}^{(2)}=e/4a^2$.  A second
scenario, shown in Fig.~\ref{fig:layers}(b), would result from the
alternation of I-V and II-IV materials, only one of which is
ferroelectric, while the other is paraelectric.  In this case,
ideal compensation requires $P_s=e/2a^2$ for the ferroelectric
component.  Intermediate cases, with the alternation of a strong
and a weak ferroelectric, are also possible.

In the remainder of this paper, we focus on the scenario sketched
in Fig.~\ref{fig:layers}(b) as realized in the SrTiO$_3$/KNbO$_3$
superlattice system, using density-functional calculations to
demonstrate the compensation mechanism proposed above.  The choice
of SrTiO$_3$ as the II-IV paraelectric component is motivated by
the fact that it is a well-studied material
\cite{vanderbilt-coissms97,rabebook} and by the common use of
SrTiO$_3$ as a substrate for growth of thin perovskite films.  With
this in mind, we assume coherent epitaxy of our superlattices on
SrTiO$_3$, so that the in-plane lattice constants are constrained
to the experimental value $a_0=$ 3.905\,\AA\ of bulk SrTiO$_3$.
As explained above, ideal compensation would require that the
ferroelectric I-V component should have a spontaneous polarization
of $P_{\rm s}=\mathrm{e}/2a_0^2=0.525\,$C/m$^2$.  We have chosen
KNbO$_3$ for the I-V component because it provides a reasonable
match to this value.  While bulk KNbO$_3$ is a rhombohedral
ferroelectric with polarization along (111) at low temperature, its
lattice constant is nearly 3\%
larger than that of SrTiO$_3$.  The calculations of Di\'eguez {\it
et al.} \cite{dieguez05} have shown that SrTiO$_3$ should become a
tetragonal ferroelectric with polarization along (001) when
compressed in-plane to fit to the SrTiO$_3$ lattice constant.
Moreover, those same calculations indicated that the polarization
of KNbO$_3$ would be $\sim$0.45\,C/m$^2$ under these conditions,
which is fairly close to the target value.  KNbO$_3$ also has the
advantage of being a commonly used and very well-studied I-V
perovskite\cite{vanderbilt-coissms97,rabebook}.

The open-source plane-wave density-functional code
\textsc{pwscf}\cite{pwscf} was used for the calculations, with the
local-density approximation to exchange and
correlation\cite{perdewzunger}
and use of ultrasoft pseudopotentials\cite{ultrasoft}.  Because we
have no reason to expect the appearance of in-plane polarization
components in this system (see above), we have assumed tetragonal
P4\textit{mm} symmetry throughout.

We first calculate the theoretically optimized value of the lattice
constant $a_0$ for bulk SrTiO$_3$ using an $8\times 8\times8$
k-point grid and a plane-wave energy cutoff of 30 Rydberg,
obtaining a value of value of 3.849 \AA.
This leads to a theoretical ideal matching condition of $P_{\rm
s}$=0.54\,C/m$^2$. Using the same k-point grid and cutoff to study
bulk KNbO$_3$ in P4\textit{mm} symmetry with its in-plane lattice
constant constrained to this $a_0$, we calculate its spontaneous
polarization to be 0.42 C/m$^2$.  These results are in good
agreement with previous work.\cite{dieguez05}

Of primary importance is whether the compensation of the bound
charge is sufficient to maintain the insulating nature of the
system as the supercell size is increased.  We consider
superlattices consisting of $n$ unit cells of SrTiO$_3$ alternating
with $n$ unit cells of KNbO$_3$, so that the supercell contains
10$n$ atoms.  Figure \ref{fig:supercell} illustrates the case of
$n$=2.  Relaxations of the multilayered supercells are performed
for values of $n$ ranging from 1 to 5. The plane-wave energy cutoff
is 30 Rydberg in all cases, and the k-point grid is $8\times
8\times M$ with $M=4$ for $n=1$ and $M=2$ for $n\ge2$.

\begin{figure}
 \includegraphics[width=2.9in]{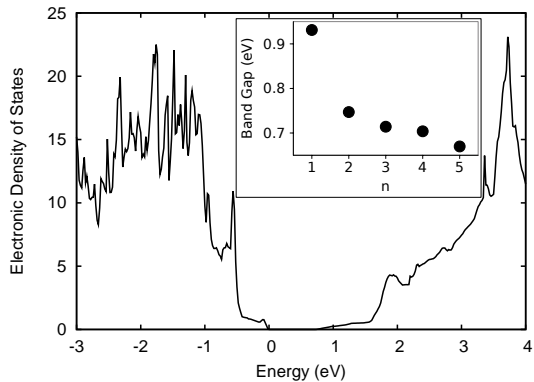}
\caption{\label{fig:4doscomb} Calculated electronic density of
states for a supercell with $n=2$ (two SrTiO$_3$ layers
and two KNbO$_3$ layers). Inset shows the the gradual reduction
of band gap for $n$SrTiO$_3$/$n$KNbO$_3$ supercells as $n$
increases.} \end{figure}

We find that the system remains insulating in all these cases.  The
density of states for $n=2$ is shown in Fig.~\ref{fig:4doscomb},
showing a clear gap between valence and conduction band states.
There is, however, a gradual closing of the band gap as $n$ is
increased, as may be expected from the imperfect charge
compensation. The calculated band gap as a function
of the supercell size is shown in
the inset to Fig.~\ref{fig:4doscomb}.  The results for $n\ge 2$
suggest a nearly linear decrease in band gap with increasing $n$.
This reduction is rather modest; a simple linear extrapolation
suggests that the system would not become metallic until
$n\simeq32$.

In order to understand the resulting local polarizations in these
multilayered systems, it is useful to start from a simple model and
test whether its predictions are borne out by a more detailed
analysis.  Due to the imperfect compensation in the system, if one
starts by assembling regions of spontaneously polarized KNbO$_3$
alternating with regions of unpolarized SrTiO$_3$, one finds planar
charge densities $\pm\sigma$ at the interfaces, with $\sigma=P_{\rm
s}-e/2a^2=-0.12$\,C/m$^2$.  Assuming that the screening of this
charge can be treated using a linear dielectric analysis and that
the thicknesses of the two constituents are approximately the same,
one finds a screened charge density of $\sigma_{\rm scr}=
2\sigma/(\epsilon_1+\epsilon_2)$, where $\epsilon_{1,2}$ are the
dielectric constants of KNbO$_3$ and SrTiO$_3$ respectively.  The
screened polarizations then become $P_{\rm KNbO_3}= P_s-\sigma
(\epsilon_1-1)/(\epsilon_1+\epsilon_2)$ and $P_{\rm
SrTiO_3}=\sigma(\epsilon_2-1)/(\epsilon_1+\epsilon_2)$.  Since
$\sigma<0$, we expect $P_{\rm KNbO_3}$ to be enhanced slightly
beyond its spontaneous value, and $P_{\rm SrTiO_3}$ to have a small
value of the opposite sign.

To investigate the correctness of this picture, we perform more
precise calculations of the local polarization profile.  An
accurate method for obtaining layer polarizations was introduced in
Ref.~~\onlinecite{wu06}, where a one-dimensional Wannier
analysis\cite{marzarivand} was employed. To do this, the overlap
matrices $M_{mn}^{(k)}=\left<u_{mk}\right|\left.u_{n,k+b}\right>$
between neighboring k-points along strings in the $\hat{z}$
direction are computed, where $u_{mk}$ is the periodic part of the
Bloch function $\psi_{mk}$. The singular value decomposition
$M=V\Sigma W^\dagger$ is used to obtain $\tilde{M} = UW^\dagger$,
which is exactly unitary.  The eigenvalues $\lambda_m$ of the
product $\Lambda=\prod \tilde{M}^{(k)}$ of these matrices along the
k-point string yield the Wannier centers as $z_{m} = (-c/2\pi) \,
\textrm{Im ln} \lambda_m$.  These Wannier centers form into
``sheets'' of charge that are localized in the growth direction but
delocalized in the plane.  For each layer $j$, we define the layer
center $z_{0,j}$ to be the average position of the ions associated
with that layer. The ``intralayer polarization'' is then given by
\begin{equation}\label{eq:lp} p^{\rm il}_j=\frac{1}{S}\sum_{\tau\in
j}Q_\tau R_{\tau z} - \frac{2e}{S} \sum_{m\in j} \bar z_m,
\end{equation}
where $S=a^2$ is the basal cell area, $Q_\tau$ is the core charge
of ion $\tau$ belonging to layer $j$,
$R_{\tau z}$ is the position of the ion measured
relative to $z_{0,j}$, and $\bar{z}_m$ is the position of the
Wannier center $z_m$ relative to $z_{0,j}$ after a $k_x$ and $k_y$
average over wavevector strings.

As long as each layer is neutral, the total polarization of the
supercell can be obtained just as $(\sum_j p^{\rm il}_j)/c$ where
$c$ is the supercell lattice constant, because the dipole moment
of a neutral layer is independent of origin.  For a supercell
containing charged layers like those associated with an I-V
material, this sum needs to be modified to $(\sum_j p^{\rm il}_j +
\sigma_jz_{0,j})/c$ in order to remain meaningful as a total
polarization. Here $\sigma_j=\pm e/S$ for a NbO$_2$ or KO layer
respectively, or $0$ for a TiO$_2$ or SrO layer.  However, the
second term does not take the form of a sum over layer
contributions.  To cast the sum into this form, we define a
layer-offcentering polarization $p_j^{\rm lo}$ that reflects the
displacement of the layer charge $\sigma_j$ from the average
position of its neighbors according to
\begin{equation}
p^{\rm lo}_j = \sigma_j  \left( \, \frac{1}{2}\,z_{0,j} 
  - \frac{1}{4}\,z_{0,j-1} - \frac{1}{4}\,z_{0,j+1} \,\right).
\end{equation}
In an extended region of I-V layers, the sum $\sum_j p_j^{\rm lo}$
counts each charged layer once and only once, but this counting is
violated at the interfaces with the II-IV layers.  For example, at
a NbO$_2$/SrO interface, we have not accounted for a charge density
of $e/4S$ in the last NbO$_2$ layer, and we have incorrectly
assigned a charge of $-e/4S$ to the first SrO layer.  The missing
charge is equivalent to a charge density of $e/2S$ located midway
between the $z_{0,j}$ values of these two layers.  Similarly, a
TiO$_2$/KO interface is assigned a charge $-e/2S$ located halfway
between these layers.  Thus, the total polarization of the
supercell can finally be written as $(\sum_j [ p_j^{\rm
il}+p_j^{\rm lo}] + \sum_\mu z^{\rm int}_\mu \sigma^{\rm
int}_\mu)/c$, where $z_\mu^{\rm int}$ and $\sigma_\mu^{\rm int}$
are the positions and charges of the extra interface charges.

The charges $\sigma^{\rm int}_\mu$ are, of course, nothing other
than the compositional polar discontinuity charges $\sigma_{\rm
comp}$ discussed earlier.  Having accounted for these, we are left
with total layer polarizations $p_j=p_j^{\rm il}+p_j^{\rm lo}$
associated with each layer, which thus provide a layer-by-layer
picture of the polarization in this type of system. To convert
these into local polarizations $P_j$ having units of charge per
unit area, we let $P_j=p_j/c_j$ where $c_j =
(z_{0,j+1}-z_{0,j-1})/2$. 

\begin{figure}
\includegraphics[width=2.6in]{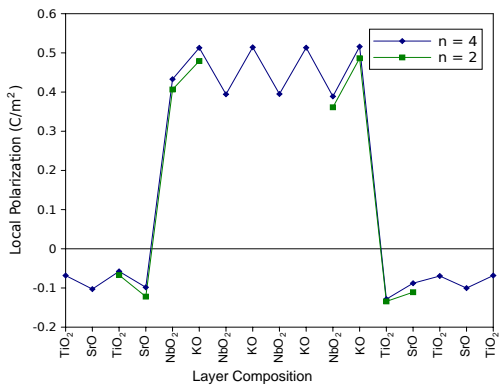}
\caption{\label{fig:n4n2wcpol} (Color online) Layer-by-layer
local polarization profile of a supercell with $n=4$ (four
SrTiO$_3$ units and four KNbO$_3$ units), with corresponding
results for $n=2$ also indicated near each of the interfaces.}
\end{figure}

The results for $n=4$ are shown in Fig.~\ref{fig:n4n2wcpol}. These
are very indicative of the results obtained for all the supercell
sizes examined. The results for $n=2$ centered on each interface
are also shown on the corresponding segments in the figure,
allowing the similarity in the behavior of the layer polarization
to be clearly seen.  A saw-tooth--like variation about an
approximately constant value is evident in examining layers
belonging to a single ABO$_3$ constituent. The polarization of the
AO layer is larger in magnitude than the BO$_2$ in both the I-V and
II-IV materials in the system, as has been observed elsewhere
\cite{wu06}.  However, there is a noticeable modification of the
TiO$_2$ layer that is adjacent to KO at the TiO$_2$/KO
heterointerface; the polarization of this layer is enhanced, so
that the saw-tooth behavior is broken.

We find the average polarization deep in the KNbO$_3$ region to be
about 0.45\,C/m$^2$, while in the SrTiO$_3$ region it is about
$-$0.08\,C/m$^2$.  Recalling that $P_{\rm s}$ of bulk KNbO$_3$ at
this in-plane lattice constant was found to be 0.42,C/m$^2$, we
find that our earlier expectations for the behavior of the
polarization profile, that $P_{\rm KNbO_3}$ should be enhanced
relative to its bulk $P_{\rm s}$ while $P_{\rm SrTiO_3}$ should be
small and of opposite sign, are clearly confirmed.

In summary, our calculations on perovskite superlattices composed
of II-IV and I-V components have shown that sharp local
polarization discontinuities can be stabilized at the charged
heterointerfaces that are intrinsic to such superlattices.
Focusing on the SrTiO$_3$/KNbO$_3$ system, our calculations show
that while the charge compensation is not perfect, the superlattice
period can be increased to rather large dimensions before the
remaining uncompensated bound charges drive the system metallic.
The same principles should apply to II-IV/III-III perovskite
superlattices.  By suitable choice of materials, and/or by tuning
the polarization via the application of epitaxial
strain, which can have a strong effect on the spontaneous
polarization \cite{dieguez05}, it should be possible to obtain even
better compensation.  The resulting superlattices have strongly
broken inversion symmetry, and may be of interest for
piezoelectric, pyroelectric, non-linear optical, and other
applications.  Finally, we have also demonstrated how a
layer-by-layer Wannier analysis may be applied to a perovskite
system in which non-neutral layers are present, as an important
step towards a more complete understanding of the local behavior of
multilayered superlattices.

\begin{acknowledgments}
The work was supported by ONR grant N00014-05-1-0054.  The
calculations were performed using the CPD cluster at The College of
William and Mary, Williamsburg, VA.  
\end{acknowledgments}

\bibliography{knosto}
\end{document}